\documentclass{article}
\usepackage{epsf}
\usepackage{complex-systems}
\usepackage{graphicx}

\ifx\pdfoutput\@undefined\usepackage[usenames,dvips]{color}
\else\usepackage[usenames,dvipsnames]{color}
\IfFileExists{pdfcolmk.sty}{\usepackage{pdfcolmk}}{} 
\fi
\usepackage[plainpages=false,pdfpagelabels,pagebackref=false,naturalnames=true,hyperindex=true,pdftitle={On the Dynamic Qualitative Behaviour of Universal Computation},pdfauthor={Hector Zenil}]{hyperref}
\hypersetup{colorlinks=true,
urlcolor=Cerulean,linkcolor=BrickRed,citecolor=RoyalBlue,a4paper,
  pdfpagemode=None,
  pdfstartview=FitH}
\usepackage[all]{hypcap}

\newenvironment{myenumerate}{
\begin{enumerate}
  \setlength{\itemsep}{1pt}
  \setlength{\parskip}{0pt}
  \setlength{\parsep}{0pt}}{\end{enumerate}
}

\begin{document}

\title{On the Dynamic Qualitative Behaviour of Universal Computation}
\author{\authname{Hector Zenil%
\thanks{hector.zenil-chavez@malix.univ-paris1.fr}}
\\[2pt] 
\authadd{Department of Computer Science / Kroto Research Institute}\\ 
\authadd{University of Sheffield, Regent Court, 211 Portobello, S1 4DP, UK}\\
}

\date{}

\maketitle

\begin{abstract}
We explore the possible connections between the dynamic behaviour of a system and Turing universality in terms of the system's ability to (effectively) transmit and manipulate information. Some arguments will be provided using a defined compression-based transition coefficient which quantifies the sensitivity of a system to being programmed. In the same spirit, a list of conjectures concerning the ability of Busy Beaver Turing machines to perform universal computation will be formulated. The main working hypothesis is that universality is deeply connected to the qualitative behaviour of a system, particularly to its ability to react to external stimulus--as it needs to be programmed--and to its capacity for transmitting this information.\medskip

\noindent \textsc{Classification:} 89.75.-k, 05.10.-a, 89.20.-a, 89.20.Ff
\end{abstract}

\section{Introduction}

In \cite{zenil} an investigation of the dynamic properties of computing machines using a  general lossless compression approach led to reasonable classifications of elementary Cellular Automata (CA) and other systems, classifications corresponding to Wolfram's four classes of behaviour \cite{wolfram}. In the spirit of other analytical concepts for scale predictability (for example, Lyapunov exponents), but employing different means, this compression-based method also led to the definition of a phase transition coefficient as a way of detecting a system's (in)stability vis-\`a-vis its initial conditions and of measuring its dynamic ability to carry information. A conjecture relating the magnitude of this coefficient and the capability and efficiency with which a system performs universal computation was introduced. In this paper the conjecture is developed further, with some additional arguments. 

In \cite{contest}, a related conjecture concerning other kinds of simply defined programs was presented, establishing that all Busy Beaver Turing machines may be capable of universal computation, as they seem to share some of the informational and complex properties of systems capable of universal computational behaviour. The conjecture will be regarded in light of algorithmic complexity, particularly of Bennett's logical depth \cite{bennett}, and will be reconnected to the first conjecture via the dynamical properties of these machines through the compression-based phase transition coefficient.

 Some definitions of concepts to be discussed either as foundations of these possible new connections, or as evidence for making such claims will be introduced first. The investigation is meant to be an exploration of empirical observations through quantitative measures which attempt to capture qualitative properties of the dynamic behaviour of systems capable of computational universality.

\subsection{Preliminaries}

Proof-of-universality results for simple programs have traditionally relied on localized structures (or ``particles''), as distinguished from relatively uniform regions.  This means that a measure of entropy of a system will tend to be below its theoretical maximum. At the same time, however, this ``particle-like'' behaviour is, and must in principle be unpredictable for the system to reach computational universality. 

Stephen Wolfram has classified all the one-dimensional nearest neighborhood CA into four classes \cite{wolfram}: (i) Class 1: ordered behaviour; (ii) Class 2: periodic behaviour; (iii) Class 3: random or chaotic behaviour; (iv) Class 4: complex behaviour. The first two are totally predictable. Random CA are unpredictable. Somewhere in between, in the transition from periodic to chaotic, complex, interesting behaviour can occur. 

One of Wolfram's open problems \cite{wolfram2} in cellular automata, for example, is the question of the computational universality of a system belonging to Class 3 (random-looking, such as rule 30) for which an entropy measure remains near its maximum at every time step, and which is unlikely to show any ``particle-like'' behaviour. The question is whether such a ``hot system'' can carry information and be programmed. The techniques to prove such a system universal may require methods different from those hitherto used for systems in which structures can be distinguished and which can therefore be made to carry information through them. The common belief is that these kinds of systems may be powerful enough but are just too complicated, perhaps even impossible to program. The encoding required to deal with the sophistication of a class III rule cellular automaton would itself probably have to possess the sophistication of a computationally universal system. This brings us to Wolfram's PCE, which states that almost all processes that are not obviously simple can be viewed as computations of equivalent sophistication (\cite{wolfram}, pp. 5 and 716-717).

\subsection{The behaviour of simple programs}

In 1970, Conway invented an automaton, which was popularised by Gardner \cite{gardner} and was known as the Game of Life. It was proved that Life was capable of universal computation \cite{berlekamp}. The proof of universality uses what in the jargon of CA are known as gliders, glider guns, and eaters, that is, structures to carry and manipulate information through the system (by combining such emergent propagating structures one can simulate logic gates and circuits). 

Langton's ant \cite{langton} is a two-dimensional Turing machine with 2 symbols and 4 states following a set of very simple rules \footnote{(a) If the machine head is on a black square, it turns 90 degrees right and moves forward one unit. (b) If the head is on a white square, it turns 90 degrees left and moves forward one unit. (c) When the head leaves a square, it prints the opposite colour.}. In \cite{gajardo}, a very simple construction is presented which proves that Langton's ant is also capable of universal computation. 

But an exhaustive exploration of one-dimensional elementary CA (that by most standards would be considered the simplest possible CA) unlike any previous system that has been constructed, was undertaken in \cite{wolfram}. The rule with number 110 (and equivalent rules: 124, 137 and 193) in Wolfram's numbering scheme, presenting the characteristic ``particle-like'' structures, turned out to be capable of universal computation \cite{wolfram,cook}. Rule 110 can be set up with initial configurations that have signals transmitted in the form of collisions of ``particle-like'' dynamical structures, simulating a variant of a tag system, another rewriting system capable of universal computation. 

The proofs of universality of all these systems imply that their dynamics are unpredictable. The notion of universality implies the existence of undecidable problems related to most questions concerning these machines. Questions related to these simple dynamical systems cannot therefore be algorithmically answered. From which it follows that undecidability is a measure of the unpredictability of a system associated with its dynamical behaviour.

\subsection{Quantitative measures of qualitative behaviour}

\noindent\textsc{Definition 1 \cite{kolmogorov,chaitin,levin}.} $K_U(s) = \min\{|p|, U(p)=s\}$ where $|p|$ is the length of $p$ measured in bits.\\

A measure of complexity is derived by combining the algorithmic complexity describing a system and the time it takes to produce a string. Bennett's concept of Logical Depth \cite{bennett,bennett2} is a complexity measure capturing the structure of a string defined by the time that a Turing machine takes to reproduce the said string from its (near) shortest description. Formally,\\

\noindent\textsc{Definition 2.} A string's logical depth D is given by $D(s)$=$\min\{t(p) : (|p|<|p_i|) \vee U(p) = s\}$\footnote{Bennett provides a careful elaboration \cite{bennett} of the notion of logical depth taking into account near-shortest programs as well as the shortest ones.}\\

According to this measure, the longer it takes, the more complex the string. Complex objects are therefore those which can be seen as ``containing internal evidence of a nontrivial causal history.''

\section{Compression-based phase transition coefficient}

A measure based on the change of the asymptotic direction of the size of the compressed evolutions of a system for different initial configurations (following a proposed Gray-code enumeration of initial configurations) was presented in \cite{zenil}. It gauged the resiliency or sensitivity of a system vis-\`a-vis its initial conditions. This phase transition coefficient led to an interesting characterisation and classification of systems, which when applied to elementary CA, yielded exactly Wolfram's four classes of systems behaviour, with no human intervention. The coefficient works by compressing the changes of the different evolutions through time, normalised by evolution space, and it is rooted in the concept of algorithmic complexity, being an upper bound of the algorithmic complexity of a string. The more compressed a string, the less algorithmically complex.

 Let the characteristic exponent $c_n^t$ be defined as the mean of the absolute values of the differences of the compressed lengths of the outputs of the system $M$ running over the initial segment of initial conditions $i_j$ with $j = \{1, \ldots, n\}$ following the numbering scheme devised in \cite{zenil} based on a Gray-code optimal enumeration scheme, running for $t$ steps in intervals of $n$. Formally,\\

\noindent\textsc{Definition 3.} $c_n^t = |C(|M_t(i_1)) - C(|M_t(i_2))|+\ldots+|C(|M_t(i_{n-1})) - C(|M_t(i_n))|/t(n-1)$.\\

\noindent\textsc{Definition 4.} Let $C$ denote the transition coefficient of a system $U$ defined as $C(U) = f^\prime(S_c)$, the derivative of the line that fits the sequence $S_c$ by finding the least-squares as described in \cite{zenil} with $S_c = S(c_t^n)$ for a fixed $n$ and $t$.\\

The value $C(U)$, based on the phase transition coefficient, is a stable indicator of the degree of the qualitative dynamical change of a system $U$. The larger the derivative, the greater the change. According to $C$, rule numbers such as 0 and 30 appear close to each other both because they remain the same despite the change of initial conditions, and because their evolution cannot be perturbed. The measure indicates that rules like rule 0 or rule 30 are also incapable of or inefficient at transmitting any information, given that they do not react to changes in the input of the system. Odd as it may seem, this is because there is no change in the qualitative behaviour of these CA when feeding them with different inputs, regardless of how different the inputs may be--rule 0 remains entirely blank while 30 remains mostly random-looking, with no apparent emergent coherent propagating structures (other than the regular and linear pattern on one of the sides). 

On the other hand, rules such as rule 122 and rule 89 appear next to each other as the most sensitive to initial conditions, because as the investigation proves, they are both highly sensitive to initial conditions and present phase transitions which dramatically change their qualitative behaviour when starting from one or another initial configuration. This means that rules 122 and 89 can be more successfully used to transmit information from the input to the output.

\subsection{Connecting dynamic behaviour and Turing universality}

Evidently if a system is completely predictable and therefore dynamically trivial, it is decidable, and therefore not Turing universal. Rule 110 should therefore not be very predictable according to the phase transition measure, but at the same time we can expect it to be versatile enough to produce the variety needed to behave as a universal. Rule 110 is one rule about which my own phase transition classification says that, despite showing some sensitivity, it also shows some stability. Which means that one can say with some degree of certainty how it will look (and behave) for certain steps and certain initial configurations, unlike those at the top. 

\begin{quotation}
This is acknowledged by Wolfram himself when discussing rule 54 ( \cite{wolfram} page 697): `It could be that if one went just a little further in looking at initial conditions one would see more complicated behaviour. And it could be that even the structures shown above can be combined to produce all the richness that is needed for universality. But it could also be that whatever one does rule 54 will always in the end just show purely repetitive or nested behaviour--which cannot on its own support universality.''
\end{quotation}

For every CA rule, there is a definite (often undecidable) answer to the question whether or not it is capable of universal computation (or in reachability terms, whether a CA will evolve into a certain configuration). The question only makes sense if the evolution of a CA depends on its initial configuration. No rule can be universal that fixes the initial configuration once and for all (there would be no way to input an instruction and carry out an arbitrary computation). 

An obvious feature of universal systems is that they need to be capable of carrying information by reflecting changes made to the input and transmitted to the output. In attempting to determine whether a system is capable of reaching universal computation, one may ask whether a system is capable of some minimal versatility in the first place, and how efficiently it can transmit information. And this is what the phase transition measures--it indicates how well a system manages to respond to an input. Obviously, a system such as rule 0 or rule 255, which does not change regardless of the input, is trivially decidable. But a universal system should be capable of reaction to external manipulation (the input to the system) in order to behave as a universal system, that is, to be capable of simulating and reaching the output of any other universal system.\\

\noindent\textsc{Conjecture 1:} Let $U$ be a machine capable of (efficient) universal behaviour. Then $C(U) > 0$.\\

 Conjecture 1 is one-way only, meaning that it states that an efficient universal system should be equipped with these dynamical properties, but the converse does not necessarily hold, given that having a large transition coefficient by no means implies that the system will behave with the freedom required for Turing universality (a case in point is rule 22, which, despite having the largest transition coefficient, seems restricted to a small number of possible evolutions).

\subsection{Evidence and discussion of a qualitative characterisation}

The conjecture is based on the following observations:
\begin{myenumerate}
\item The phase transition coefficient provides information on the ability of a system to react to external stimuli.
\item Universal systems are (efficient) information processors capable of carrying and transmitting information.
\item Trivial systems and random-looking systems are incapable of transmitting information.
\item Trivial systems have negative $C$ values, close to zero.
\item Rules such as 110, proven to be universal, and rule 54 (suspected to be universal, see \cite{wolfram} page 697) turn out to be classified next to each other, with a positive transition coefficient.
\end{myenumerate}

The capacity for universal behaviour implies that a system is capable of being programmed and is therefore reactive to external input. It is no surprise that universal systems should be capable of responding to their input and doing so succinctly, if the systems in question are efficient universal systems. If the system is incapable of reacting to any input or if the output is predictable (decidable) for any input, the system cannot be universal. 

Values for the subclass of CA referred to as elementary (the simplest one-dimensional) have been calculated and published in \cite{zenil}. We will refrain from evaluations of $C$ to avoid distracting the reader with numerical approximations that may detract from our larger goal. The aim is to propose some basics of a behavioural characterisation of computational universality.

\begin{figure}[htdp]
\centering
\scalebox{.35}{\includegraphics{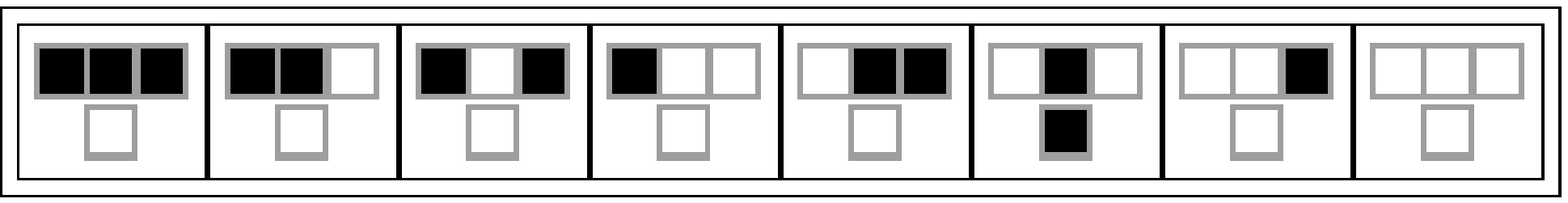}}\\
   \scalebox{.4}{\includegraphics{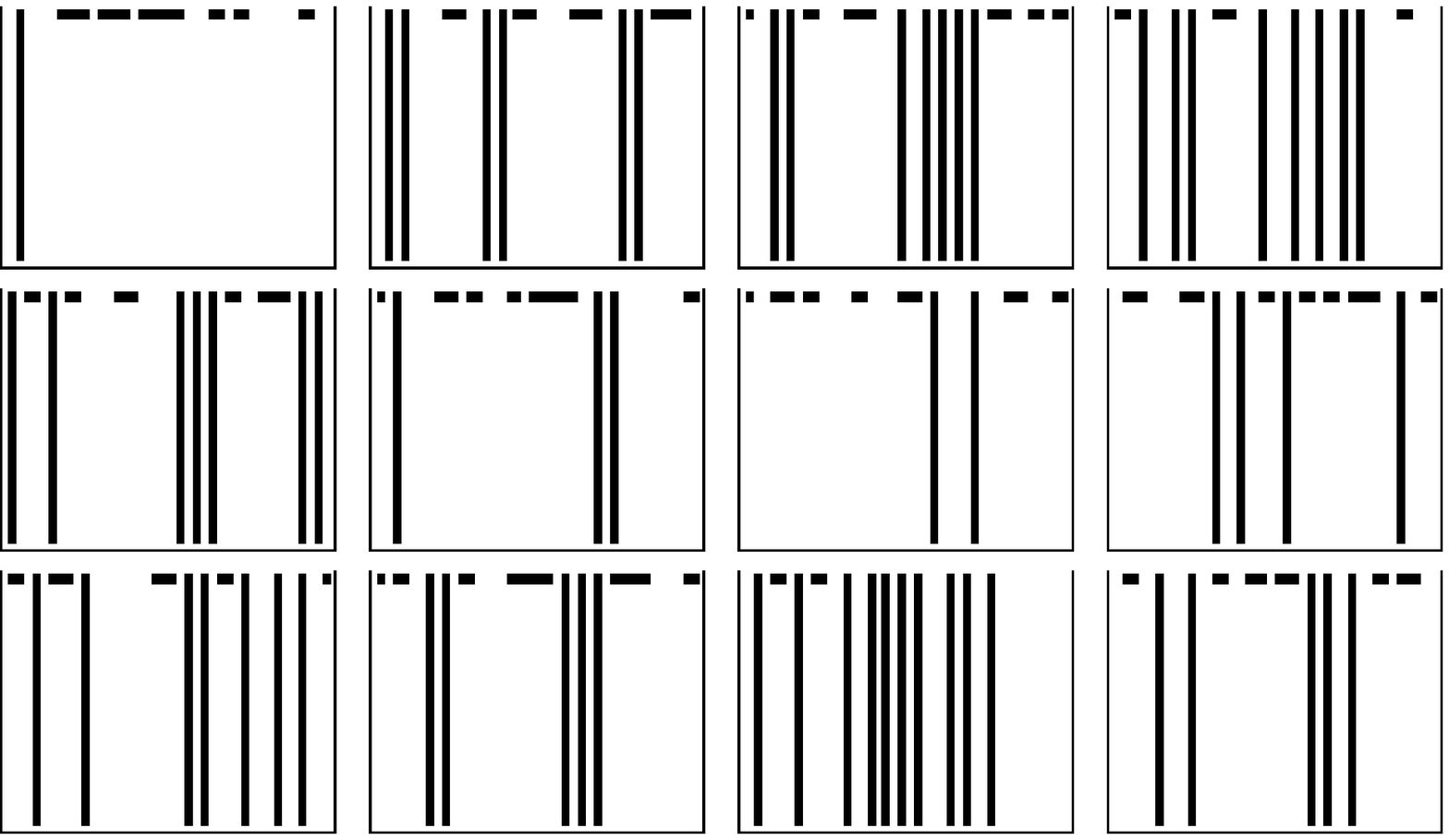}}
\caption{ECA rule 4 is a kind of program filter that only transfers bits in isolation (i.e. when its neighbors are both white). It is clear that one can perform some very limited computations with this automaton.}
\end{figure}

For example, some rules, such as rule 0, don't produce different configurations relative to variant initial configurations. No matter how one changes the initial condition, there is no way to make it produce other than what it computes for every other initial configuration. These trivial elementary CA rules are automatically ruled out, particularly the most simple among them that cannot usually be ruled out as candidates for universal behaviour given that even if they look trivial for the simplest or for certain initial configurations, they could still be capable of the necessary versatility and eventually be programmed in light of the space of all possible inputs for which they may be sensitive. The foundations of conjecture 1 and the conjecture itself are consistent with all these observations, but it is most meaningful for systems that are believed to be of great complexity but are usually not believed to be malleable enough to be programmed as universal systems, such as is the case with rule 30. If the conjecture is true, $C(U)$ may not only rule out systems which intuition strongly suggests are unable to behave as universals, but it would also indicate that random-looking systems such as rule 30 are not capable of universal computation because they are incapable of carrying information. In this sense, the measure may also be a characterisation of the practical randomness of a system in terms of efficient information transmission. 

Rule 110, however, has a positive $C$ value, meaning it is efficient at carrying information from its input through the output, and that one can actually program it to perform computations. $C$ is compatible with the fact that it has been proven that rule 110 is capable of universal computation.

\begin{figure}[htdp]
\centering
   \scalebox{.35}{\includegraphics{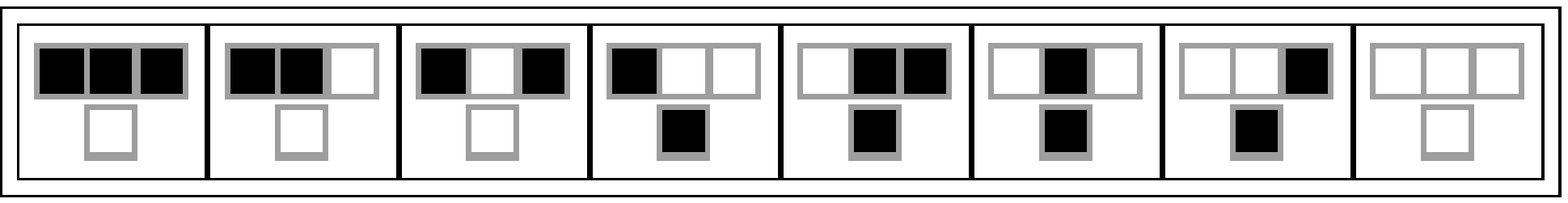}}\\
      \scalebox{.4}{\includegraphics{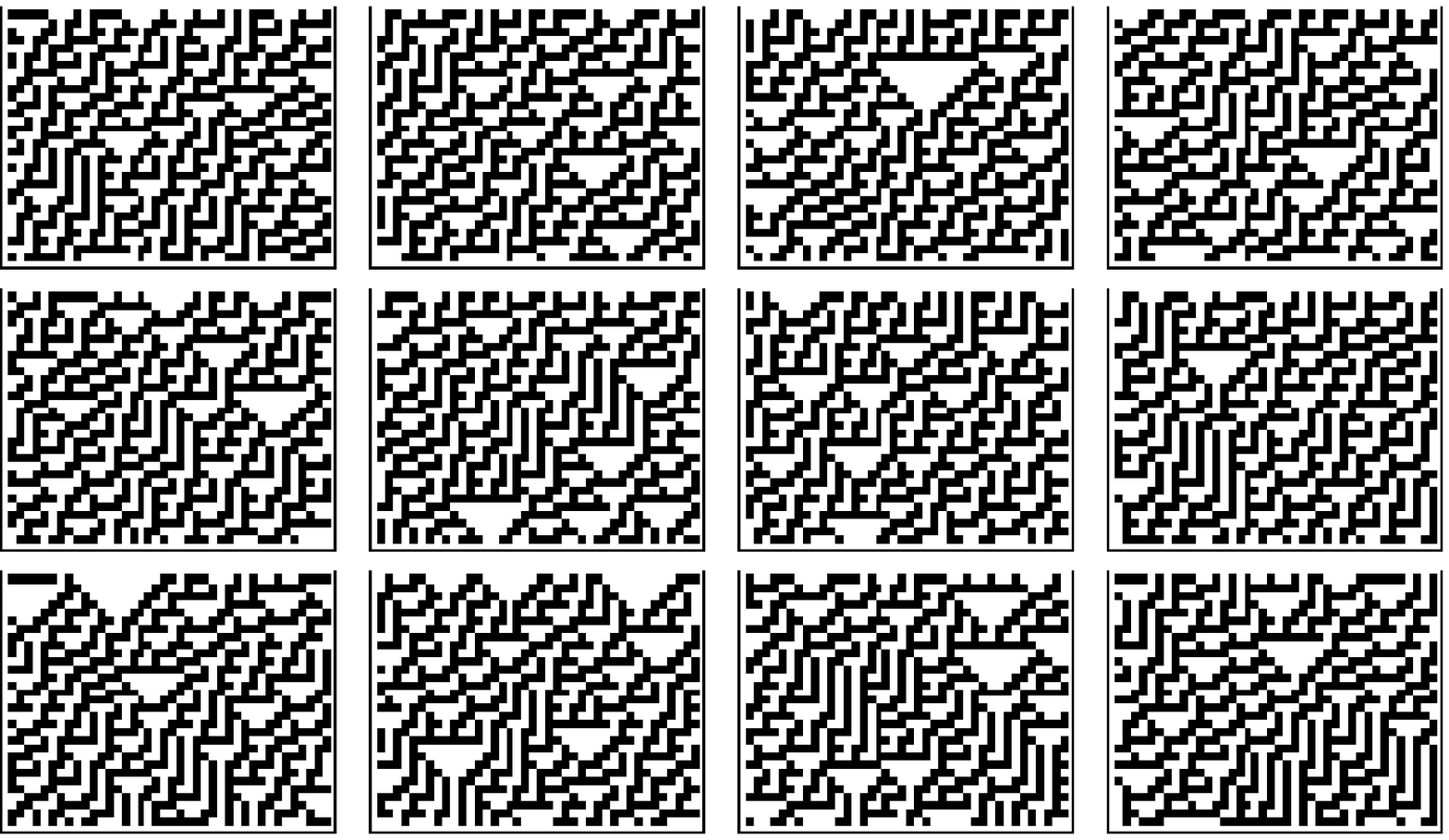}}
\caption{It is an open question whether ECA rule 30 can be programmed to perform computations. Its $C$ value is low, meaning that it is not efficient for transferring information because it always behaves in the same fashion--too randomly.}
\end{figure}

A universal computer (would therefore have a non-zero $C$ limit value. $C$ also captures some of the universal computational efficiency of the computer in that it has the advantage of capturing not only whether it is capable of reacting to the input and transferring information through its evolution, but also the rate at which it does so. So $C$ is an index of both capability in principle and ability in practice. A non-zero $C$ means that there is a way to codify a program to make the system behave (efficiently) in one fashion or another, i.e. to be programmable. Something that is not programmable cannot therefore be taken to be a computer. 

In \cite{margolus}, Margolus asserts that reversible cellular automata (RCA) can actually be used as computer models embodying discrete analogues of classical notions in physics such as space, time, locality and microscopic reversibility. He suggests that one way to show that a given rule can exhibit complicated behaviour (and eventually universality) is to show (as has been done with the Game of Life \cite{gardner} and rule 110 \cite{cook,wolfram}) that ``in the corresponding `world' it is possible to have computers" starting these automata with the appropriate initial states, with digits acting as signals moving about and interacting with each other to, for example, implement a logical gate for digital computation. 

Conjecture 1 also seems to be in agreement with Wolfram's beliefs concerning rule 30, which according to his Principle of Computational Equivalence (PCE) \cite{wolfram} may be computationally universal and still be impossible to control so as to be able to perform a computation (something that Wolfram has himself suggested \cite{wolfram}). 

RCA are interesting because they allow information to propagate, and in some sense they can be thought of as perfect computers--indeed in the sense that matters to us. If one starts an RCA from a non-uniformly random initial state, the RCA evolves, but because it cannot get simpler than its initial condition (for the same reason given for the random state) it can only get more complicated, producing a computational history that is reversible and can only lead to an increase in entropy.

\section{On the possible computational power of Busy Beaver machines}

\subsection{Busy Beaver machines}
\label{beaver}

Rado also \cite{rado} studies the behaviour of a special kind of one-tape $n$-state deterministic Turing machine, one that starts with a blank tape, writes more non-blank symbols than any other $n$-state Turing machine, and halts. 

\noindent\textsc{Notation:} We denote by $(n,2)$ the class (or space) of all $n$-state 2-symbol Turing machines (with the halting state not included among the $n$ states).

\noindent\textsc{Definition 5.} \cite{rado} If $\sigma_T$ is the number of 1s on the tape of a Turing machine $T$ upon halting, then:  $\sum(n)=\max{\{\sigma_T : T\in(n,2) \normalsize{\textbf{ }T(n)\textbf{ }halts}\}}$. If $t_T$ is the number of steps that a machine $T$ takes upon halting, then $S(n)=\max{\{t_T : T\in(n,2) \normalsize{\textbf{ }T(n)\textbf{ }halts}\}}$.

$\sum(n)$ and $S(n)$ as defined in 1 and 2 are noncomputable functions by reduction to the halting problem. Yet values are known for $(n,2)$ with $n \leq 4$.

 The Busy Beaver problem lies at the heart of what may be seen as a paradox, for while a Busy Beaver machine of $n$ states can be thought of as having maximal sophistication vis-\`a-vis all $n$ state Turing machines as regards the number of steps and printed symbols, Busy Beaver machines can be extremely easily defined. The definition of Busy Beaver machines describes an infinite set of Turing machines characterised by a particular behaviour--the attribute of printing more non-blank symbols on the tape before halting, or having the longest runtime among all Turing machines of the same size (number of states).

 Bennett's logical depth measure is relevant in characterising the complexity of an $n$-state Busy Beaver machine both in terms of size (fixed among all $n$-state machines) and in terms of the behaviour that characterises this type of machine, because it follows from Rado's definitions and Bennett's concept of logical depth that Busy Beavers are the deepest machines provided that they are the ones with the longest history producing a string.

 Yet a Busy Beaver is required to halt. When running for the longest time or writing the largest number of non-blank symbols, $bb(n)$ has to be clever enough to make wise use of its resources and still save a rule to halt. These facts may suggest the following conjectures, also in connection with the dynamic behaviour of a set of simply described machines with universal behaviour.\\

\noindent\textsc{Conjecture 2:}
\begin{myenumerate}
\item (strong version): For all $n>2$, $bb(n)$ is capable of universal computation.
\item (sparse version): For some $n$, $bb(n)$ is capable of universal computation.
\item (weak version): For all $n>2$, $bb(n)$ is capable of (weak) universal computation.
\item (weakest version): For some $n$, $bb(n)$ is capable of (weak) universal computation.
\end{myenumerate}

It is known that among all 2-state 2-symbol Turing machines none can be universal. Remember, however, that $bb(n)$ as defined by Rado \cite{rado}, is a Turing machine with $n$ states plus a special halting state. So $bb(2)$ is actually a 3-state 2-symbol machine in which one state is specially reserved for halting only. By letting $bb(n)$ be a weak universal machine, one allows initial tape configurations other than those filled with just a single symbol (usually called a blank tape, but blankness is a symbol in itself), but with initial configurations simple enough so that one can guarantee that the computation is not performed before it is given already computed in the input encoding. In other words, $bb(n)$ is allowed (in the conjecture versions 2.3 and 2.4) to start either from a periodic tape configuration or an infinite sequence of the type accepted by a regular $\omega$-automaton \cite{wolfgang}.

\subsection{Discussion of the characterisation}

If any version of the conjectures excepting conjecture 2.4 is true, the characterisation would define a countable infinite set of universal Turing machines. Their proof may provide an interesting framework and a possible path to take for proving a whole set of Turing machines to be capable of universal computation on the basis of their common dynamical properties.

 Because halting machines that always halt cannot be capable of unbounded computation, and therefore of universal Turing behaviour, among the analytical tools necessary to demonstrate the universality of (any) of these systems are proofs that Busy Beavers are capable of avoiding the halting state. If one proves that Busy Beavers always halt, that would amount to proving that they cannot be universal. But to disprove conjectures 2.1 to 2.3 one can simply prove that at least one Busy Beaver is not capable of a halting configuration, and a study of this type is likely to be simplified for bb(3) or bb(4), for which Busy Beaver functions are known and are Turing machines small enough to be subjected to a thorough and potentially fruitful investigation in this regard. The investigation of the behaviour of Busy Beaver machines for other than blank tape initial configurations indicates that these machines are capable of non-trivial behaviour for other than the simplest initial configuration (as intuition would suggest, given that if they behave in a sophisticated fashion for the simplest initial condition, they may be expected to continue doing so for more complicated ones). In a future paper we will explore the specific behaviour of these machines.

The truth of the conjectures may not seem intuitively evident to all researchers, given that it is possible that these machines are only concerned with producing the largest numbers by using all resources at hand, regardless of whether they do so intelligently. However, the requirement to halt is, from our point of view, a suggestion that the machine has to use its resources intelligently enough in order to keep doing its job while saving a special configuration for the halting state.

 Despite the conclusion that conjecture 2.4 would imply, namely that the property of being a Busy Beaver machine is not a characterisation of the computational power of this easily describable set of countable infinite machines, among the intuitions suggesting the truth of one of these conjectures is that it is easier to find a machine capable of halting and performing unbounded computations for a Turing machine if the machine already halts after performing a sophisticated calculation than it is to find a machine showing sophisticated behaviour whose previous characteristic was simply to halt. This claim can actually be quantified, given that the number of Turing machines that halt after $t=n$ for increasing values of $n$ decreases exponentially \cite{calude,delahaye}. In other words, if a machine capable of halting is chosen by chance, there is an exponentially increasing chance of finding that it will halt sooner rather than later, meaning that most of these machines will behave trivially because they won't have enough time to do anything interesting before halting.
 
We have no positive proof of any version of these conjectures and much more work remains to be done on the dynamical behaviour of these systems. But conjectures 1 and 2 lead us to:\\
 
 \noindent\textsc{Conjecture 3:} $C(bb(n))>0$.

\section{Concluding remarks}

The first conjecture relates computational universality to the capacity of a computational system to transfer information from the input to the output and reflect the changes in the evolution of the system when starting out from different initial configurations. We established that the property of having a large phase transition coefficient seems necessary. On the other hand, a universal system seems to be capable of manifesting an abundance of possible evolutions and reacting to different initial configurations in order to (efficiently) behave universally. 

A second conjecture concerning the possible universality of a kind of well-defined infinite set of abstract Busy Beaver Turing machines was introduced--also in terms of a version of a measure of complexity related to algorithmic complexity and the dynamic behaviour of these machines having a particular common characterisation. The third conjecture relates conjectures 1 and 2.

 These conjectures will be the subject of further study in a paper to follow this one. We would like to see the conjectures proved or disproved, but underlying the conjectures are many other interesting questions relating to the size, behaviour and complexity of computing machines. It would be interesting, for example, to find out whether there is a polynomial (or exponential) trade-off between program size and the concept of simulating a process.

\end{document}